# GPTutor: a ChatGPT-powered programming tool for code explanation

Eason Chen[1], Ray Huang[2], Han-Shin Chen[3], Yuen-Hsien Tseng[1] and Liang-Yi Li[1]

[1] National Taiwan Normal University, Taipei, Taiwan
[2] KryptoCamp, Taipei, Taiwan
[3] University of Toronto, Toronto, Canada
`eason.tw.chen@gmail.com`

**Abstract.** Learning new programming skills requires tailored guidance. With the emergence of advanced Natural Language Generation models like the ChatGPT API, there is now a possibility of creating a convenient and personalized tutoring system with AI for computer science education. This paper presents GPTutor, a ChatGPT-powered programming tool, which is a Visual Studio Code extension using the ChatGPT API to provide programming code explanations. By integrating Visual Studio Code API, GPTutor can comprehensively analyze the provided code by referencing the relevant source codes. As a result, GPTutor can use designed prompts to explain the selected code with a pop-up message. GPTutor is now published at the Visual Studio Code Extension Marketplace, and its source code is openly accessible on GitHub. Preliminary evaluation indicates that GPTutor delivers the most concise and accurate explanations compared to vanilla ChatGPT and GitHub Copilot. Moreover, the feedback from students and teachers indicated that GPTutor is user-friendly and can explain given codes satisfactorily. Finally, we discuss possible future research directions for GPTutor. This includes enhancing its performance and personalization via further prompt programming, as well as evaluating the effectiveness of GPTutor with real users.

**Keywords:** ChatGPT, Tutoring System, Developer Tool, Prompt Engineering, Natural Language Generation.

## 1 Introduction

Lately, there has been a rise in the need for skilled programmers, and as a result, many individuals are opting to learn coding and pursue lucrative software-related careers. At school, students are crowded in programming courses [1]. Moreover, the gap between learning and practical application requires students to continue learning after entering the workforce. For example, in 2020, 42% of beginner-level technology workers joined the US job market via the Coding Boot Camp [2]. Because of the strong demand for coding education, there is a shortage of teachers, which makes it difficult to provide personalized learning in these classrooms. Some students may feel frustrated. While self-studying and using Google to find solutions to problems can be helpful, there are times when students may require assistance when reading documents or examples for an unfamiliar programming language. Furthermore, it is especially challenging when novice people onboarding a new job and need to catch up by reading others' codes [3].

The code could include domain-specific business logics, which might be unfamiliar to them, and may be uncomment, poorly maintained, or even unclean.

This paper presents GPTutor as a remedy to relieve programmers from aforementioned issues. GPTutor is a plugin for Visual Studio Code that uses ChatGPT to provide detailed explanations of source code. With GPTutor, students can conveniently receive personalized explanations for coding problems they encounter. Additionally, those seeking to learn a new programming language can use GPTutor to understand example code. Finally, new employees needing to quickly familiarize themselves with a codebase can use GPTutor to gain insights into the business logic behind each line of code.

In sum, the main contributions of this paper are:

1. We developed GPTutor, a Visual Studio Code extension that utilizes the OpenAI ChatGPT API to provide detailed explanations of the given source code.

2. We demonstrated and explained why GPTutor surpasses other code explain applications, such as vanilla ChatGPT or GitHub Copilot, by advanced prompt designs.

3. We discussed potential applications, limitations, and future research directions on programming code explain applications like GPTutor.

## 2 Background

### 2.1 Natural Language Generation

Natural Language Generation (NLG) is a subfield of artificial intelligence (AI) that uses computer algorithms to produce human-like language output from the given input [4]. NLG aims to generate coherent and contextually appropriate language indistinguishable from human-writing language.

NLG applications may appear to provide intelligent responses to given questions, but in reality, they just guess next words based on the vast amount of data they read [5]. For example, in Figure 1, if the NLG model uses the Artificial Intelligence in Education Conference websites as its training data and receives the prompt input "International Conference on Artificial Intelligence in". In that case, the NLG model may deem "Education" as a more possible follow-up after the given input than other words. As a result, the NLG model will complete the word with "Education" then continued to generate possible follow-up texts such as "will take place July 3-7, 2023 in Tokyo". The model may also produce results such as "July 27-31, 2022 in Durham" or even a fictitious outcome like "July 20-24, 1969 on the Moon".

By providing additional contextual information in the prompt, the likelihood of the desired text being generated increases. Figure 1 demonstrates this phenomenon. When we include the prompt "The 24th" to the beginning of the prompt input, the model will be more inclined to generate "July 3-7, 2023 in Tokyo" as output since the website stated that the 24$^{th}$ AIED is held at 2023 in Tokyo. The technique of designing proper prompts to get the desired output is known as prompt programming [6].

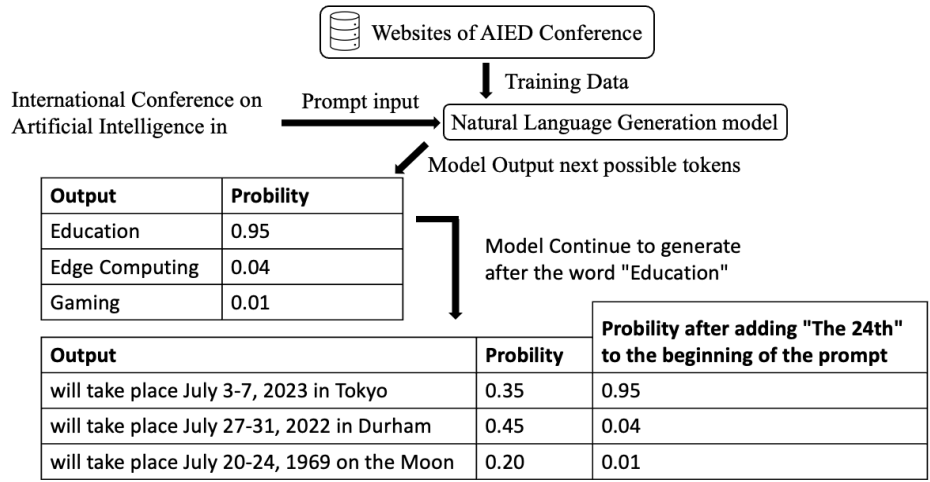

**Fig. 1.** Example of probability on generating different outputs during NLG process.

### 2.2 Using NLG for Programming Code Explanation

We could use prompt programming to employ large language models, such as GPT-3, as a tutor to answer question based on the context [4]. For example, if the NLG model was trained with lots of document about programming code and its comments/explanations, the model will be able to explain the given code like the example in Figure 2.

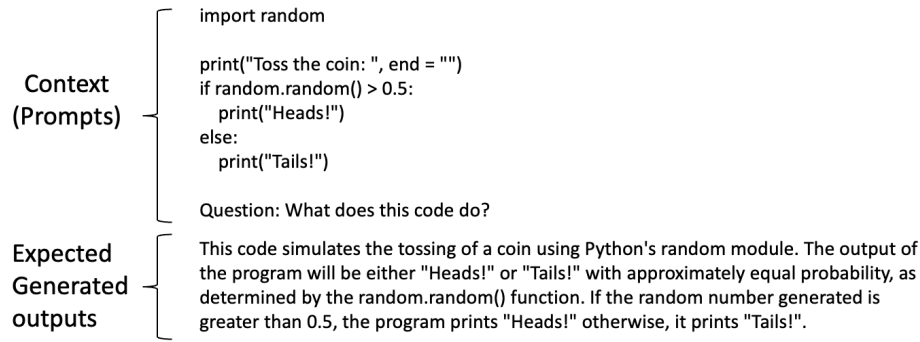

**Fig. 2.** Example of input and output on using NLG model as a code explainer.

Many existing applications, such as GPT-3, ChatGPT, and GitHub Copilot, can perform the NLG explanation as shown above in Figure 2. Nevertheless, these applications still have three main limitations, as presented in Figure 3.

First, existing NLG code explainers are superficial, as they can only offer insights based on the code present on the current file. Consequently, they may overlook or speculate domain logics behind the function. This issue becomes particularly noteworthy when analyzing code with object-oriented design that imports objects from other files.

Second, existing NLG code explainers tend to offer excessive, irrelevant, or even fictitious information. For instance, if a user asking on a line of code with GitHub Copilot, it may explain the entire code from top to bottom, which is often unnecessary.

Lastly, existing NLG code explainers may not be up to date. For example, ChatGPT was only trained with data until 2021 and, therefore, may perform well with popular libraries which had a lot of training data at that time. However, it may not provide a satisfactory explanation when dealing with new, unpopular, or private libraries.

GPTutor was developed to surpass the aforementioned limitations, as shown in Figure 3. It offers the most concise and accurate explanations. Additionally, it can provide a comprehensive analysis of the provided code by examining the function's source code.

```python
#attendeeManager.py
import os
from pymongo import MongoClient
from uuid import uuid4

class AttendeeManager:
    def __init__(self):
        client = MongoClient(os.getenv("MONGODB_URI", ""))
        conference_name = os.getenv("CONF_TITLE", "")
        self.mongo_col = client[conference_name]["attendees"]

    def add_attendee(self, email, name=None, id=None, voucher=None):
        if id == None: id = uuid4()
        attendee = {"name": name, "id": id, "email": email, "voucher": voucher}
        self.mongo_col.insert_one(attendee)
```

```python
#main.py
from attendeeManager import AttendeeManager

attendeeManager = AttendeeManager()
attendeeManager.add_attendee("john@gmail.com", "John Doe")
```

**ChatGPT** Mar 23 Version (pasting main.py and asking for why using *.add_attendee*)

The code you provided is creating an instance of the AttendeeManager class and then adding an attendee to it using the add_attendee() method. The reason why the .add_attendee() method is used is because it is a method defined within the AttendeeManager class that allows you to add an attendee to the list of attendees managed by the AttendeeManager. By calling this method with the appropriate arguments (in this case, the attendee's email address and name), the method will create a new Attendee object with the provided information and add it to the list of attendees. In summary, the .add_attendee() method is used to add new attendees to an AttendeeManager object.

Correct
Irrelevant
Fictitious (Wrong)

**GitHub Copilot Labs** v0.11.742 (Use "Ask Copilot: Explain Code" for *.add_attendee*)

Here is the explanation for the code above:
1. The first line is the import statement, which is used to import the library.
2. The second line is the API key. You can get your API key from the AttendeeManager page.
3. The third line is the API secret. You can get your API secret from the AttendeeManager page.
4. The fourth line is the AttendeeManager object.
5. The fifth line is the add_attendee method, which is used to add a new attendee to your AttendeeManager account.
6. The first parameter of the add_attendee method is the email address of the attendee.
7. The second parameter of the add_attendee method is the name of the attendee.
8. The add_attendee method returns the attendee_id of the new attendee.

**GPTutor** with *gpt-3.5-turbo* at 2023/03/05 (select *.add_attendee* and request explanation)

The code above is adding a new attendee to the MongoDB database. The add_attendee method takes in the email and name of the attendee, and also generates a unique ID (using uuid4()) and an optional voucher code (if available). In the example above, the email is "john@gmail.com" and the name is "John Doe", so a new attendee document will be created in the database with these fields. This method call is essentially populating the database with attendee information.

Only GPTutor Find out

**Fig. 3.** Example Code and the comparison of the explanation from ChatGPT, GitHub Copilot, and the GPTutor.

## 3  Implementation of GPTutor

In this section, we will first describe how we built the GPTutor Extension with Visual Studio Code API. Then, we discuss how we enhance its' performance in ChatGPT API.

### 3.1  Building GPTutor as a Visual Studio Code Extension

We built GPTutor in the Visual Studio Code extension development environment in Typescript. During the initial setup, the extension will ask users to provide their OpenAI API key, which will be stored in the extension's global state.

Then, when users request an explanation of code through the GPTutor extension by command or hot key, the extension will perform the following steps:

1. Use the "editor.document.languageId" API to determine the language of the file.
2. Use the "editor.document.getText" API to obtain the code for the current file.
3. If the cursor is positioned on a function, the GPTutor will additionally use the "editor.action.revealDefinition" API to retrieve the source code behind the function.

### 3.2  Getting answer by ChatGPT API with prompt programming

Using the data obtained from the above steps, the GPTutor will create the prompt shown in Figure 4 for the *gpt-3.5-turbo* model via the OpenAI API, which was just released on March 1, 2023. We tried several prompts and found the following formatted in Figure 4 yielded the most favorable results.

```typescript
let result = await openai.createChatCompletion({
  model: "gpt-3.5-turbo",
  messages: [
    {
      role: "system",
      content: `You are a helpful coding tutor master in ${language}.`,
    },
    {
      role: "user",
      content: `The following is the source code of the library of
        ${selectedFunctionName}:\n${sourceCodeOfSelectedFunction}\n
        The following is the ${language} code:\n${currentCode}\n
        Question: why use ${selectedText} at ${textAtLineOfCursor}
        in the ${language} code above?`,
    },
  ],
});
let explain = result.data.choices[0].message?.content;
```

**Fig. 4.** The prompts GPTutor used to feed into the gpt-3.5-turbo model.

## 4  Current Results

GPTutor has been published on the Visual Studio Code Extension Marketplace at https://marketplace.visualstudio.com/items?itemName=gptutor.gptutor, and its source code is openly accessible at https://github.com/GPTutor/gptutor-extension. Preliminary user interview with students, programming teachers, and coding boot camp tutors indicated that GPTutor is user-friendly and can explain any given code satisfactorily. GPTutor especially impresses users with its remarkable ability to incorporate relevant source codes behind functions into prompts to provide a thorough explanation.

## 5 Discussion and Future Works

### 5.1 Enhance performance and personalization by prompt programming

GPTutor's superior performance compared to other similar applications can be attributed to its use of more relevant code in its prompts. This enables the NLG model to provide more desirable answers. We will continue to enhance GPTutor's performance by optimizing prompts. One possible way is by using heuristic search to identify relevant code in the code base. Then, after transforming the codes into many possible prompts, GPTutor could provide various explanations [7] to find users preference and then offer them personalized explanations and a better user experience.

### 5.2 Evaluate the effectiveness of using GPTutor in the real world

We will investigate the impact of GPTutor on students' comprehension of programming by observing how they interact with it to complete programming assignments. To assess the effectiveness of GPTutor, we will collaborate with coding course lecturers and utilize the Between-Subjects Design and the Interrupted Time Series Analysis to measure the relationship between the student grades and the frequency of the use of GPTutor.

## 6 Conclusion

We created GPTutor, an extension for Visual Studio Code that leverages ChatGPT to provide programming code explanations. GPTutor collects relevant code and utilizes the OpenAI ChatGPT API to explain the chosen code. Comparisons indicate that GPTutor delivers the most concise and accurate explanations compared to Vanilla ChatGPT and GitHub Copilot. We believe that GPTutor can enhance computer science education and offer each student a convenient and personalized learning experience in the future.

**Acknowledgement.** This work was supported by the Ministry of Science and Technology of Taiwan (R.O.C.) under Grants 109-2410-H-003-123-MY3 and 110-2511-H-003-031-MY2. We thank the KryptoCamp for the use cases and preliminary evaluation.